\documentclass[preprint,proceedings]{rmaa}

\SetYear{2016}
\SetConfTitle{XV Latin American Regional IAU Meeting LARIM}

\title{Astroparticle Techniques: Simulating cosmic rays induced background radiation on aircrafts}

\author{H. Asorey\altaffilmark{1}, L. A. N\'u\~nez\altaffilmark{2,3}, C. Y. P\'erez-Arias\altaffilmark{2}, S. Pinilla\altaffilmark{2}, F. Qui\~nonez\altaffilmark{2} and M. S\'uarez-Dur\'an\altaffilmark{2}.
}

  \altaffiltext{1}{Laboratorio Detecci\'on de Part\'iculas y Radiaci\'on, Centro At\'omico Bariloche \& Instituto Balseiro, Bariloche, Argentina}
  \altaffiltext{2}{Escuela de F\'isica, Universidad Industrial de Santander, Bucaramanga, Colombia.}

  \altaffiltext{3}{Departamento de F\'isica, Universidad de Los Andes, M\'erida, Venezuela}

\suppressfulladdresses

\listofauthors{H. Asorey}
\listofauthors{L. A. N\'{u}\~{n}ez}
\listofauthors{C. Y. P\'erez Arias}
\listofauthors{S. Pinilla}
\listofauthors{F. Qui\~{n}onez}
\listofauthors{M. Su\'arez-Dur\'an}
\indexauthor{Asorey, H.}
\indexauthor{N\'u\~nez, L.~A.}
\indexauthor{P\'erez Arias, C.~Y.}
\indexauthor{Pinilla, S.}
\indexauthor{Quinonez, F.}
\indexauthor{Su\'arez-Dur\'an, M.}

\addkeyword{Cosmic Rays}
\addkeyword{Radiation Biological effects}
\addkeyword{Aviation}
\addkeyword{Dose Rate}

\begin{document}
% Typeset article header
\maketitle 

\boldabstract{Incident cosmic ray fluxes over flying aircrafts are compared with those in Bucaramanga, Colombia and very significant differences are observed for proton and neutron fluxes. We also obtained that major contributions in the deposited energy by Cherenkov photons on blood plasma is in the UV-C band.}

\textit{Cosmic Rays} (particles and nuclei with energies from $10^{5}$\,eV to $10^{20}$\,eV) enter into the  atmosphere generating a cascade of particles impinging on aircrafts flying between 10\,km to 12\,km. It has been found that at these altitude airplanes are exposed to cosmic ray radiation levels up to two order of magnitude higher than at sea level \citep{PinillaAsoreyNunez2015}.  

Integrated particle flux and its modulation are carefully calculated and corrected by considering local atmospheric profiles and dynamic geomagnetic conditions at a constant altitude of $11$\,km\,\citep{AsoreyEtal2015B}. Subsequently, these results are piped into a GEANT4/GATE simulation platform\,\citep{JanEtal2011} to model the interaction of high energy particles with a spherical blood plasma phantom of $0.1$\,m$^{3}$.

Hight energy secondary particle flux at flight level directly (for charged particles) and indirectly (trough, e.g., pair creation) generate Cherenkov photons in the medium. The Cherenkov energy spectra and the corresponding deposited energy in blood plasma are estimated for five flight trajectories: BOG-BUE, BUE-MAD, JNB-SYD, JFK-HND and SAO-JNB. These flights were selected due different geomagnetic features they cross, such as the Arctic oval or the South Atlantic Anomaly. % Bogot\'a-Buenos Aires (BOG-BUE); Buenos Aires-Madrid (BUE-MAD), Johannesburgo-Sidney (JNB-SYD), New York-Tokyo (JFK-HND), S\~ao Paolo-Johannesburgo (SAO-JNB). 

Every $30$\,minutes the flux is calculated with the corresponding local atmospheric profile and secular geomagnetic conditions and this flux value is assumed constant during the next $30$\,minutes track.  

When compared with a reference point (Bucaramanga, Colombia, 965 m.a.s.l.), very significant differences on the relative flux ($\Delta N = (N_{\mathrm{Route}} - N_{\mathrm{BGA}})/N_{\mathrm{BGA}}$) for photons, protons and neutrons are observed (see Table \ref{TableBMGFlight}). Our calculations show that the major deposited energy contribution in blood comes from Cherenkov photons in the UV-C $100$\,nm$-280$\,nm band. With these calculated values for UV exposure, it seems that it is possible to induce some damage at cellular level\,\citep{PradaEtal2016}.
\begin{table}[!t]
  \includegraphics[width=0.9\columnwidth]{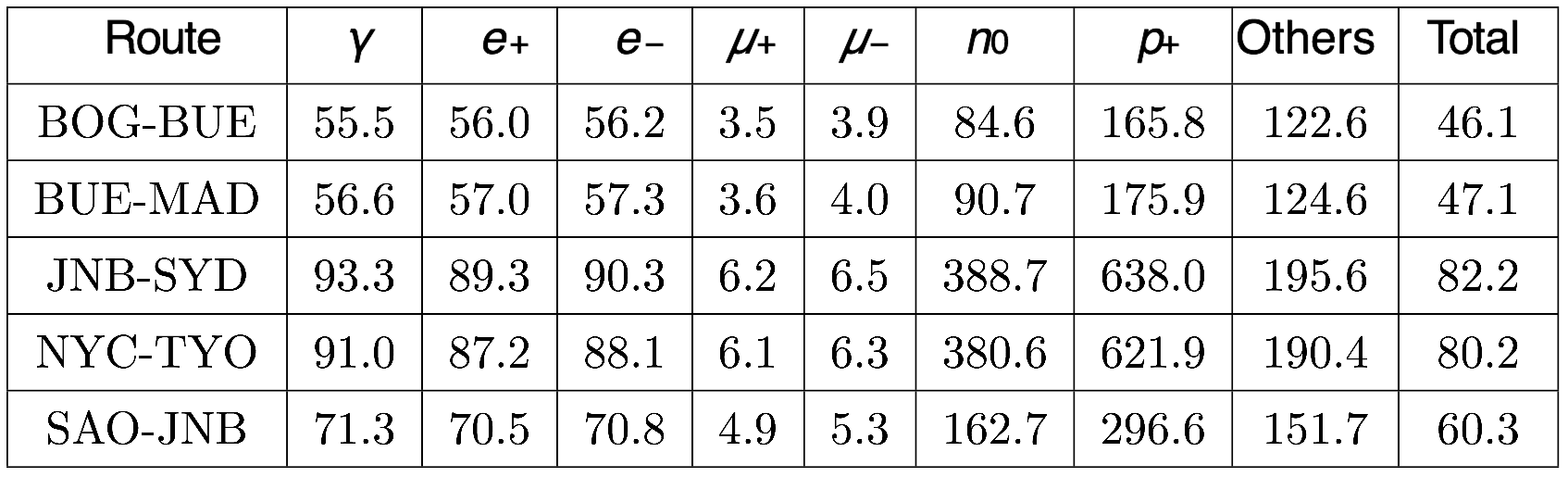}
	\caption{Number of secondaries at flight level relative to Bucaramanga.}
	\label{TableBMGFlight}
\end{table}

\begin{figure}
\begin{center}
\includegraphics[width=0.8\columnwidth]{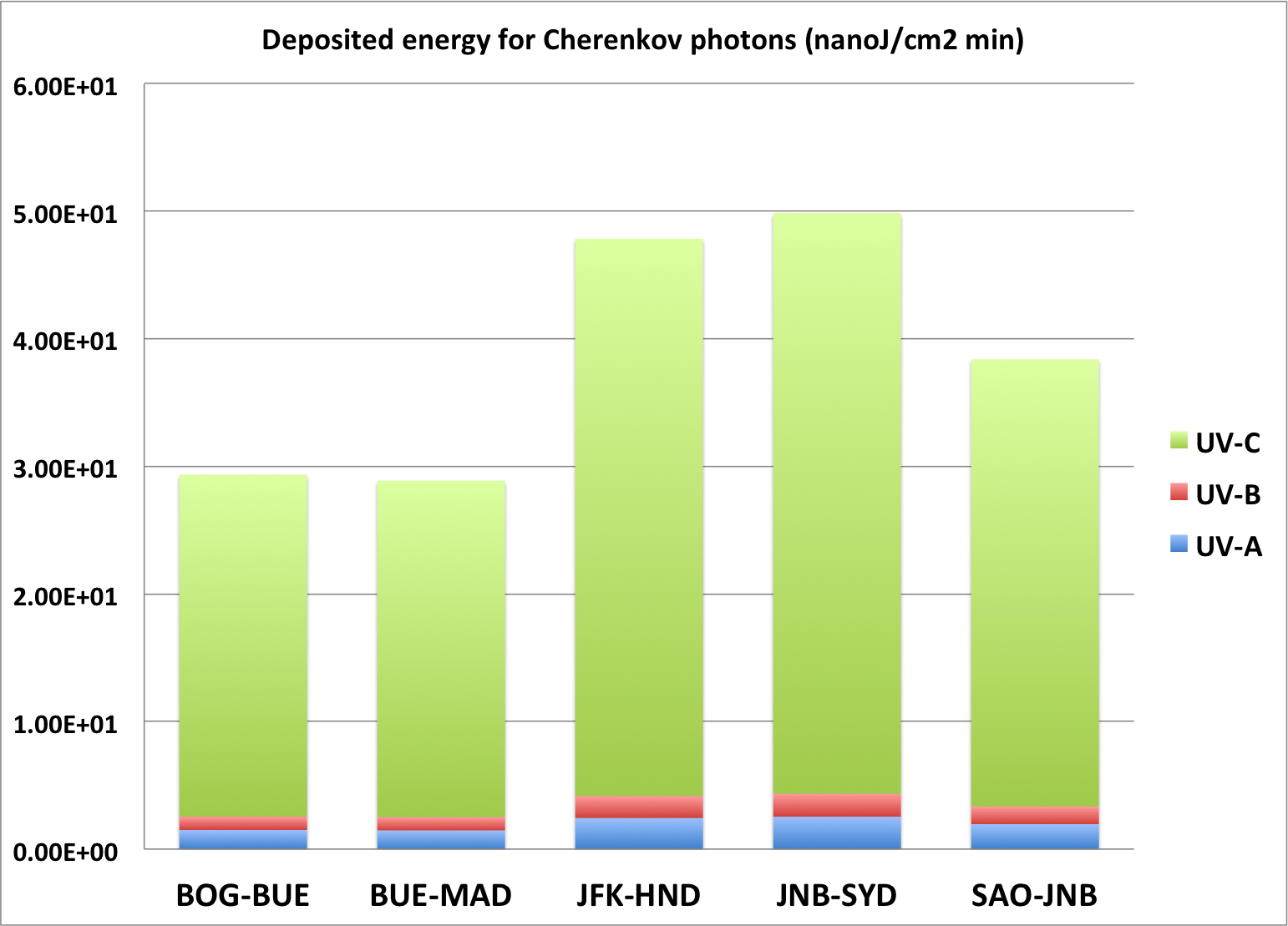}
	\caption{Deposited energies in plasma by Cherenkov photons in the standard ultraviolet (UV) bands: UVA ($315$\,nm$-400$\,nm), UVB ($280$\,nm$-315$\,nm) and UVC ($100$\,nm$-280$\,nm).}
\label{EnergiaFotonesCherenkovEng}
\end{center}
\end{figure}


\begin{thebibliography}
\expandafter\ifx\csname natexlab\endcsname\relax\def\natexlab#1{#1}\fi
\expandafter\ifx\csname href\endcsname\relax
  \def\href#1#2{}\fi
\expandafter\ifx\csname urllinklabel\endcsname\relax
  \def\urllinklabel{[LINK]}\fi
\expandafter\ifx\csname adsurllinklabel\endcsname\relax
  \def\adsurllinklabel{[ADS]}\fi

\bibitem[{Asorey {et~al.}(2015)Asorey, Dasso, N\'u{\~n}ez, P\'erez,
  Sarmiento-Cano, Su\'arez-Dur\'an}]{AsoreyEtal2015B}
Asorey, H., Dasso, S., N\'u{\~n}ez, L., P\'erez, Y., Sarmiento-Cano, C.,
\&  Su\'arez-Dur\'an, M. 2015, in 34th
  International Cosmic Ray Conference, PoS(ICRC2015), 142


\bibitem[{OpenGate Collaboration (2011)}]{JanEtal2011}
OpenGate Collaboration 2011, Physics in Medicine and Biology, 56, 881


\bibitem[{Pinilla {et~al.}(2015)Pinilla, Asorey, \&
  N{\'u}{\~n}ez}]{PinillaAsoreyNunez2015}
Pinilla, S., Asorey, H., \& N{\'u}{\~n}ez, L. 2015, Nuclear and Particle
  Physics Proceedings, 267-269, 418


\bibitem[{Prada-Medina {et~al.}(2016)Prada-Medina, Aristizabal-Tessmer,
  Quintero-Ruiz, Serment-Guerrero, \& Fuentes}]{PradaEtal2016}
Prada-Medina, C.A., Aristizabal-Tessmer, E.T., Quintero-Ruiz, N.,
Serment-Guerrero, J., \& Fuentes, J.L. 2016, International Jour. of
  Radiation Biology, 1


\end{thebibliography}
\end{document}